\documentclass[conference]{IEEEtran}
\IEEEoverridecommandlockouts

\usepackage[T1]{fontenc}
\usepackage{cite}
\usepackage{amsmath,amssymb,amsfonts}
\usepackage{textcomp}
\usepackage{enumerate}
\usepackage{array}
\usepackage{stfloats}
\usepackage{url}
\usepackage{verbatim}
\usepackage{graphicx}
\usepackage{subcaption}
\usepackage{diagbox}
\usepackage{booktabs}
\usepackage{multirow}
\usepackage{float}
\usepackage{bm}
\usepackage{hyperref}
\usepackage[table]{xcolor}
\graphicspath{{./figures/}}
\def\BibTeX{{\rm B\kern-.05em{\sc i\kern-.025em b}\kern-.08em
T\kern-.1667em\lower.7ex\hbox{E}\kern-.125emX}}
\begin{document}

    \title{Towards Objective Estimation of Physical Functioning Status in Breast-Cancer Patients using Real-World Data
    \thanks{This work was supported by the EU Horizon 2020 framework under Grant Agreement No. 965231}
    \thanks{Vasileios Papapanagiotou is also an Assistant Professor with the Department of Medicine of Karolinska Institutet, Sweden (IMPACT group)}
    }

    \author{
        \IEEEauthorblockN{
            Alexandros Papadopoulos,
            Ioannis Sarafis, \\
            Vasileios Papapanagiotou
            and Anastasios Delopoulos
        }
        \IEEEauthorblockA{
            \textit{School of Electrical and Computer Engineering} \\
            \textit{Aristotle University of Thessaloniki, Greece}\\
            \{palexang, iosarafis, papapavi, antelopo\}@ece.auth.gr
        }
        \and
        \IEEEauthorblockN{
            Marius Stensland,
            Kristin Jonsdottir,\\
            Ingrid Holsvik Dahl,
            Marie Austdal,\\
            Tone Hoel Lende,
            and Kjersti Tjensvoll
        }
        \IEEEauthorblockA{
            \textit{Stavanger University Hospital, Norway}\\
            \{marius.stensland, kristin.jonsdottir, ingrid.holsvik.dahl,\\
            marie.austdal, tone.hoel.lende, kjersti.tjensvoll\}@sus.no
        }
    }

    \maketitle

    \begin{abstract}
        Breast cancer is the most common invasive cancer in women, accounting for about one third of cancer cases.
        As with all cancers, its treatment may introduce side-effects
        like fatigue and
        reduced physical functioning, leading to a downward spiral
        of subsequent issues and severely affecting patient quality of life.
        On the positive side, physical functioning can
        be improved via personalized targeted interventions, tailored to the particularities of each case. However, continuous and objective monitoring of patients is a difficult task,
        owing to the limited clinical resources and the lack of appropriate tools.
        In this paper, we introduce a methodology
        for objectively estimating the physical functioning status of breast cancer patients from wearable smartwatch data.
        We show that features derived from longitudinal walking and heart rate measurements correlate significantly with physical functioning scores
        of established patient-reported outcome measures.
        In doing so, our work takes a first step towards developing novel validated
        digital biomarkers for estimating physical functioning in breast cancer patients,
        thus reducing the need for self-reported measures and costly clinical evaluations.
    \end{abstract}


    \section{Introduction}
    Health-related Quality of Life (HRQoL) is a multifaceted concept that captures
    various, often interrelated, aspects of a person’s functional status. According
    to the EORTC QLQ-C30 questionnaire \cite{fayers_quality_2002}, overall quality
    of life can be decomposed into specific domains, including physical, role,
    emotional, cognitive, and social functioning. Improving the quality of life of
    cancer patients is a primary goal of supportive care intervention. This is
    particularly crucial for \emph{Breast Cancer Patients (BCPs)}, who often face a range
    of post-treatment symptoms such as cancer-related fatigue and a general decline
    in physical functioning \cite{zomkowski_physical_2018}. In these cases, targeted
    and personalized interventions can play a key role in alleviating symptoms and
    enhancing daily well-being. Assessing a patient's physical functioning status
    is, therefore, essential for enabling such interventions.
    Currently, this assessment is carried out through clinical examination,
    while in research studies, standardized
    tools like \emph{Patient-Reported Outcome Measures (PROMs)} are also employed.
    However, these point-in-time assessments provide only a limited, snapshot-like view of a
    patient’s evolving condition, with continuous and objective assessment of
    physical functioning remaining challenging, owing to the inherent limitations
    and finite nature of clinical resources. This is where digital health tools
    can offer a promising alternative.

    The idea of using digital tools for assessing physical functioning is not new.
    The very interesting approach by Paraschiv-Ionescu et al.
    \cite{paraschiv-ionescu_unraveling_2013}, used accelerometer and gyroscope data
    from body-worn sensors to model real-life physical activity in chronic pain
    patients, introducing a methodology for capturing the temporal dynamics between
    activity and rest periods. Luo et al. \cite{luo_assessment_2020} used sensor
    data (accelerometer, barometer, galvanic skin response, temperature and
    PPG) from Everion wearables to produce daily fatigue
    predictions using machine learning. They reported interesting
    results on a cohort of 27 healthy subjects when compared with their self-
    reported scores. Other works have employed different sensors such as ECG and
    metrics like the HRV for fatigue estimation \cite{mu_intelligent_2024}.

    Apart from specific symptoms like fatigue or pain, one can estimate the overall
    physical functioning status of a person. To that end, Gupta et al.
    \cite{gupta_feasibility_2018} investigated the feasibility of using physical
    activity monitors to measure the functional status of 24 cancer patients on
    systemic therapy. They reported correlations between derived measures, such as
    mean step count, and physician-assessed functional status using the ECOG
    Performance Status. Gresham et al. \cite{gresham_wearable_2018} used
    commercially-available smartwatches by Fitbit to assess functional status in
    37 advanced cancer patients. In their feasibility study, they also found strong
    correlations between device-measured step counts and clinician-assessed
    functional status as measured by the ECOG/Karnofsky scales and NIH PROMIS.
    However, their results were limited to patients in advanced cancer stages.

    In this paper, we focus on the functional status of breast-cancer patients.
    It has been established that BCPs stand to benefit greatly from targeted and
    personalized physical activity interventions \cite{ballinger_personalized_2021,
    bender_physical_2021}. As a result, developing digital tools for the continuous
    assessment of BCPs' functional status holds much promise for improving their
    quality of life. This is one of the goals of the REBECCA project\cite{bourgos_2025_17378308}, a Horizon 2020
    research and innovation programme, whose overarching aim is to use Real-World
    Data (RWD) to support, complement and enrich clinical research and to improve
    existing clinical workflows. Spanning multiple clinical studies across 3
    countries (Norway, Spain and Sweden), REBECCA collected RWD from breast-cancer
    patients using off-the-shelf wearables and smartphones through the course of
    18 months.

    Driven by the data collected in REBECCA, here we conduct a systematic exploratory
    analysis to investigate whether objective indicators derived from 24/7
    smartwatch measurements of step counts and heart rate are informative
    regarding BCP physical functioning status.
    We demonstrate that several such indicators correlate significantly and
    consistently with self-reported physical functioning scores across
    multiple clinical visits, establishing a foundation for future predictive
    modeling. In doing so, this work provides evidence for the feasibility of
    using wearable-derived digital biomarkers as objective proxies of physical
    functioning in breast cancer patients, a necessary first step toward replacing
    or complementing costly clinical evaluations and self-reported measures.

    \section{Methodology}

    \begin{figure*}[!ht]
        \centering
        \includegraphics[width=\linewidth]{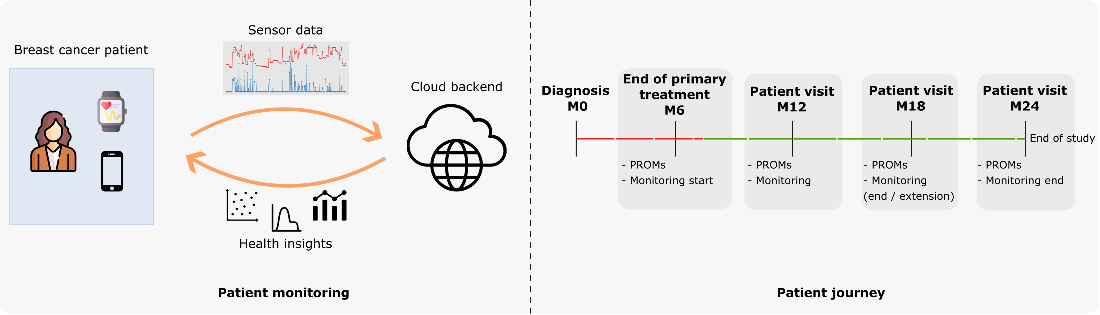}

        \caption{High-level overview of the patient journey (right) and monitoring procedure (left) during their participation in the REBECCA-CRF study}

        \label{fig:overview}
    \end{figure*}

    \subsection{The REBECCA-CRF study}

    The data used in this study were collected during the REBECCA-CRF (Cancer Related
    Fatigue) clinical study\cite{stensland2026monitoring} conducted at Stavanger University Hospital, Norway. The
    study recruited BCPs (M0 stage, $\leq$80 years) at the time of diagnosis. At
    M0, participants went through a clinical assessment, during which they were
    asked to complete PROMs, provided informed consent, and underwent a short pre-
    treatment medical exam (height, weight, blood pressure) with blood sample
    collection. Following completion of primary treatment in Month 6 (M6),
    participants were introduced to the REBECCA system, which included installation
    of the REBECCA mobile app, setup of the wearable device, and instructions on
    their use.
    Follow-up visits were scheduled after 6 and 12 months (M12 and M18 in the study timeline).
    An overview of the participant's journey is given in Fig. \ref{fig:overview}. In
    some cases, study participation was extended to 18 months, with an additional
    visit conducted accordingly at M24. At each visit, participants were clinically
    evaluated and completed PROMs, including the EORTC QLQ-C30 and SF-36
    questionnaires. The inclusion period ran from May 2022 to June 2023, leading
    eventually to 57 BCPs and 53 healthy women controls actively using REBECCA to
    contribute real-world data. Out of these, a subset of 43 participants
    (Table\ref{tab:cohort_demographics}) was ultimately used in this paper, due to
    withdrawals and data collection issues.

    \begin{table}[!h]
        \centering
        \caption{Demographic and clinical characteristics of the BCP and HC cohorts
        of the REBECCA-CRF study used in our analyses.
        Values are presented as mean$\pm$std (min-max).}
        \label{tab:cohort_demographics}
        \begin{tabular}{lcc}
            \toprule
            \textbf{} & \parbox{3cm}{\centering \textbf{BCP} \\ (n=23 participants)}      & \parbox{3cm}{\centering \textbf{HC} \\ (n=20 participants)}      \\
            \midrule
            Age & 54.5$\pm$10.9 (32-78)   & 47.0$\pm$14.6 (24-78)   \\
            BMI  & 25.5$\pm$5.1 (18.3-38.4)   & 25.3$\pm$3.5 (19.4-31.6)    \\
            EORTC PF  & 87.4$\pm$18.6 (33-100)   & 94.1$\pm$10.9 (33-100)    \\
            SF36 PF  & 87.9$\pm$19.7 (25-100)   & 94.2$\pm$8.3 (70-100)    \\
            \bottomrule
        \end{tabular}
        \caption*{\footnotesize \textit{PF: Physical Functioning, BCP: Breast Cancer Patient, HC: Health Controls}}
    \end{table}

    \subsection{Continuous monitoring}
    Study participants contributed monitoring data through the REBECCA mobile app, a
    cross-platform smartphone application that served as the central tool for data
    collection of REBECCA. The app supports continuous and unobtrusive data
    collection from both wearable devices and the sensors of the smartphone
    itself. To that end, the app integrates seamlessly with Garmin™ wearables,
    allowing the collection of continuous per-minute measurements of heart rate,
    stress level, sleep patterns, and step counts. On the smartphone side, it
    supports continuous location data collection via the smartphone’s built-in GPS
    sensor, to identify mobility patterns such as time at home, radius of gyration,
    and other metrics. Aside from passive monitoring, the app also supports self-
    reported data collection, enabling patients to upload annotated photos (e.g.,
    meals, drinks, and significant life events—positive or negative) and complete
    short in-app questionnaires inspired by standardized Patient-Reported Outcome
    Measures (PROMs) related to functional status.

    Study participants were instructed to use the REBECCA mobile app in conjunction
    with a Garmin Venu SQ smartwatch. Data were periodically transmitted from the
    smartwatch to the participant's smartphone, and then to a secure cloud-based
    server for storage and further analysis (Fig. \ref{fig:overview}).
    Participants were asked to use the system throughout their participation in the
    study (they could drop out at any time).

    \subsection{Missing data}
    Despite the goal of continuous 24/7 data collection, substantial periods of
    missing data were frequently observed. This was due to a variety of factors
    such as participants not wearing the device, whether for charging, comfort, or
    personal reasons, intermittent connectivity issues between the smartwatch and
    the smartphone, as well as incorrect device use (e.g., failing to grant
    required permissions or disabling Bluetooth). Software integration issues
    between the mobile app and the smartwatch were also a significant problem that
    was largely mitigated in subsequent system versions. In general, such
    challenges are characteristic of real-world data collection in longitudinal
    studies and are therefore expected when deploying digital health tools outside
    controlled clinical settings.

    To address this, we applied a minimum data sufficiency criterion: only
    participants with at least one complete week of valid sensor data within a two-month
    window centered on a clinical visit were included in the analysis. The two-month
    window was chosen deliberately to maximize the amount of data available for
    feature extraction, given the high prevalence of missing periods.
    Participants who did not meet this threshold for any given visit were
    excluded from the corresponding feature–ground truth pair, though they
    were retained for visits where sufficient data were available.
    This per-visit exclusion policy accounts for the fact that data
    availability varied across participants and time points, and is the
    primary reason for the reduction from the originally recruited cohort of
    57 BCPs and 53 healthy controls to the 23 BCPs and 20 healthy controls
    ultimately used in our analyses. While we cannot fully rule out selection
    bias (participants with more complete data may systematically differ from
    those with less) the per-visit rather than per-participant exclusion approach
    minimizes unnecessary data loss and reduces the risk of systematically excluding
    participants based on overall compliance alone.

    \subsection{Functional status ground-truth}
    As functional status ground-truth, we used the EORTC QLQ-C30
    \cite{fayers_quality_2002} and SF-36 \cite{treanor_methodological_2015}
    questionnaires, completed during each visit of a participant to the clinic (M6,
    M12, M18). Both include a \emph{Physical Functioning (PF)} scale that evaluates
    ability to perform daily activities, though with some differences in
    scope. The EORTC QLQ-C30 \cite{fayers_quality_2002} is a widely used
    questionnaire for evaluating quality of life in cancer patients, containing a
    PF scale of five items that estimate the patient’s ability to perform daily
    activities such as walking, climbing stairs, and engaging in physical work. The
    SF-36 \cite{treanor_methodological_2015} is a general health survey that also
    includes a PF scale, assessing the ability to carry out physical activities
    such as walking, running, and lifting heavy objects. In our experiments, we
    used the outcomes of both PF scales as ground-truth, scored on the 0–100 range
    (according to the respective questionnaire specifications), with higher scores
    indicating better physical functioning.
    Notably, scores across the two scales were found to be highly
    correlated (r=0.87, p$<$0.05), thus justifying the complementary
    use of both.

    \subsection{Physical functioning features}

\begin{figure}[!h]
        \centering
        \includegraphics[width=\linewidth]{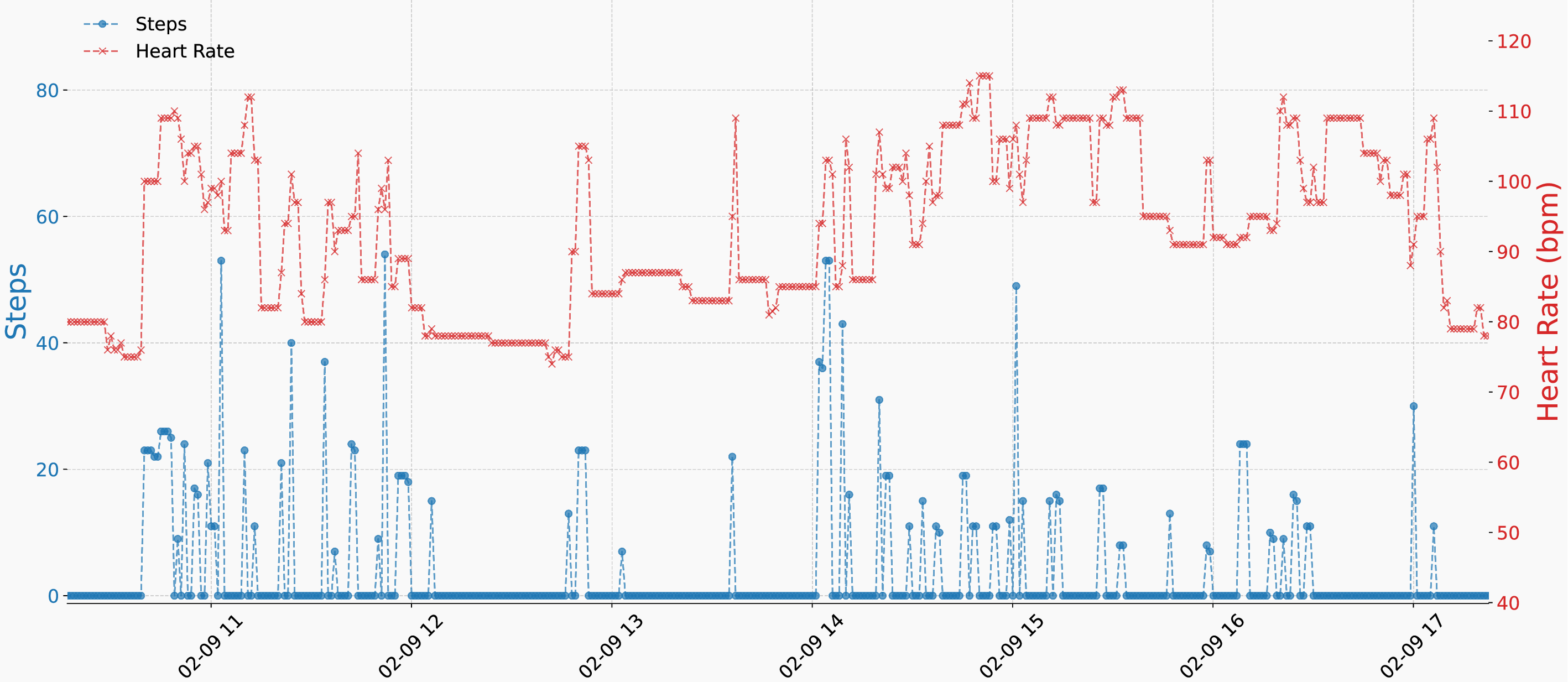}
        \caption{Example of raw heart rate and step count time series from a single
        patient over 6 hours, a small excerpt of the full two-month data window
        used in our analysis.}
        \label{fig:raw_data}
    \end{figure}

   Despite the rich data modalities captured by the REBECCA system, in this work we
   focus exclusively on step count and heart rate data from the Garmin smartwatch.
   Although the REBECCA system also collects sleep, stress, and continuous GPS recordings,
   incorporating these modalities would have further reduced the already limited
   number of eligible participants and visit-level observations, given their
   comparatively lower availability in the dataset. Stress data were also
   initially considered but were ultimately excluded due to high noise levels and
   low overall informative content. Step count and heart rate were therefore
   selected as the primary signals, offering the best balance between data
   completeness and physiological relevance to physical functioning. Specifically,
   we analyzed two time series: heart rate (HR) and step counts (steps), both
   sampled at a 1-minute resolution by the Garmin smartwatch, with heart rate
   expressed in beats per minute (bpm) and step counts representing the number of
   steps taken within each minute. An example of the raw form of these two time
   series is shown in Figure~\ref{fig:raw_data}.

    The raw heart rate time series underwent a mild pre-processing pipeline
    designed to impute missing values and reduce measurement noise. Specifically,
    linear interpolation was applied to gaps of up to 2 consecutive minutes, while
    longer missing segments were left unchanged. Heart rate values below 40 bpm or
    above 200 bpm were deemed erroneous and treated as missing. The steps time
    series was used without modification. Following pre-processing, a set of
    features (Table \ref{tab:features}) was extracted, either from each signal
    separately (average daily steps, resting heart rate, steps and heart rate
    statistics), or by analyzing them together (heart rate to step rate ratio). To
    compute the entropy features, we approximated the underlying distributions
    using binning. For steps, we applied uniform bins spanning 0 to 200 steps per
    minute with a bin width of 25. For heart rate, we used uniform bins ranging
    from 40 to 180 bpm with a width of 10. The 180 bpm upper limit was chosen as it
    reflects the age-predicted maximum heart rate of the study cohort.

\begin{table}[h]
    \centering
    \rowcolors{2}{gray!20}{white}
    \caption{Overview of the extracted physical functioning features}
    \label{tab:features}
    \begin{tabular}{c l p{0.55\linewidth}}
        \hline
        \textbf{Symbol} & \textbf{Feature} & \textbf{Description} \\
        \hline
        $HR_{rest}$      & \textit{restingHr}      & Heart rate while at rest \\
        $\bar{S}$        & \textit{avgDailySteps}  & Average daily step count within window \\
        $R_{HR/S}$         & \textit{hrToStepRatio}  & Average heart rate to steps per minute ratio \\
        $H_s$            & \textit{stepsEntropy}   & Entropy of steps per minute distribution \\
        $\gamma_{1,s}$   & \textit{stepsSkewness}  & Skewness of steps per minute distribution \\
        $\gamma_{2,s}$   & \textit{stepsKurtosis}  & Kurtosis of steps per minute distribution \\
        $H_{hr}$         & \textit{hrEntropy}      & Entropy of heart rate (bpm) distribution \\
        $\gamma_{1,hr}$  & \textit{hrSkewness}     & Skewness of heart rate (bpm) distribution \\
        $\gamma_{2,hr}$  & \textit{hrKurtosis}     & Kurtosis of heart rate (bpm) distribution \\
        \hline
    \end{tabular}
\end{table}

    \section{Results}
    In this section, we evaluate whether the extracted set of features is
    informative with respect to physical functioning. To that end, for each
    participant we defined a two-month window centered on each clinical visit. We
    then extracted the physical functioning features of Table \ref{tab:features}
    from the data of this period and associated them with EORTC-QLQ-C30 and SF-36
    physical functioning scores available from that visit. Cross-referencing the
    available PROMs with participants who had sufficient sensor data within the
    corresponding time window resulted in a dataset of 51 feature–ground truth
    pairs from 23 BCPs and 31 pairs from 20 healthy controls.

    \subsection{Univariate Statistical Analysis}

    First, we investigate whether the extracted features are
    statistically different between the BCP and HC populations. To that
    end, we perform the Mann-Whitney U test and report the results in
    Table\ref{tab:man-whitney-results}. For running the test, we computed the mean
    of each feature across visits of the same participant, so as not to violate the i.i.d assumption of the Mann-Whitney formulation.
    For completeness, we also
    report the differences between the two physical functioning scale scores
    (QLQ-C30 and SF36) across the two groups.
    We find that four features (steps entropy,
    steps kurtosis, steps skewness, and the heart rate to steps ratio) differ
    significantly between the two populations ($p < 0.05$), each with a moderate
    effect size as indicated by the rank-biserial correlation $r$.

    \begin{table}[h!]
    \centering
    \rowcolors{2}{gray!15}{white}
    \caption{Mann-Whitney U test results for the
    BCP ($n_1$=23) and
    HC ($n_2$=20)
    groups of the REBECCA-CRF study.
    Column $r$ denotes rank-biserial correlation.}
    \label{tab:man-whitney-results}
    \begin{tabular}{lcccc}
        \toprule
        \textbf{Feature} & \textbf{Median BCP} & \textbf{Median HC} & \textbf{$r$} & \textbf{p-value} \\
        \midrule
        $\gamma_{1,s}$               & 1.46 & 1.10  & 0.487  & 0.007 \\
        $R_{HR/S}$            & 3.82 & 3.30  & 0.417  & 0.020 \\
        $\gamma_{2,s}$               & 1.95 & 0.41  & 0.409  & 0.023 \\
        $H_s$                & 1.27 & 1.40  & -0.396 & 0.028 \\
        \hline
        QLQ-C30 PF & 93.3 & 100.0 & -0.526 & 0.003 \\
        SF-36 PF   & 93.8 & 98.1  & -0.335 & 0.057 \\
        \bottomrule
    \end{tabular}
\end{table}

    We can see that the BCP group exhibits significantly lower steps entropy
    compared to the HC group, suggesting more predictable step patterns. The step
    distribution skewness and kurtosis are also notably higher in this group, as
    well as the heart rate to step rate ratio, suggesting that patients expend more
    cardiovascular effort per unit of walking speed. This is consistent with
    reduced cardiorespiratory fitness or a higher physiological cost of movement
    in the patient population.

    \subsection{Correlation Analysis}

    We now shift our focus to the physical functioning status of the two groups.
    Specifically, we examine whether the proposed features are associated with
    participants' self-reported physical functioning status, as measured by the
    EORTC QLQ-C30 and SF-36 questionnaires completed at each clinical visit. The
    results of the correlation analysis for the BCP group are reported
    in Table\ref{tab:corr_analysis_results}. To account for the possibility of multiple
    data points per BCP, we conducted separate analyses for each clinical evaluation
    event (visit), so that each participant is always represented by a single data point.
     \begin{table}[h!]
        \centering
        \caption{Correlation analysis of extracted features with physical functioning ground-truths, stratified by clinical visit (Event). Values denote Pearson's $r$ followed by p-value in parentheses. Significant correlations ($p<0.05$) are marked in bold.}
        \label{tab:corr_analysis_results}

        \begin{tabular}{lcccc}
            \toprule
            \multicolumn{5}{c}{\textbf{EORTC-QLQ-C30 Physical Functioning Scale}} \\
            \midrule
            \textbf{Feature} & \textbf{M6 visit} & \textbf{M12 visit} & \textbf{M18 visit} & \textbf{M24 visit} \\
            & \small{(n=15)} & \small{(n=16)} & \small{(n=12)} & \small{(n=8)} \\
            \midrule
            $H_s$ & \textbf{0.76} (<0.01) & \textbf{0.71} (<0.01) & \textbf{0.75} (<0.01) & \textbf{0.89} (<0.01) \\
            $R_{HR/S}$ & \textbf{-0.71} (<0.01) & \textbf{-0.74} (<0.01) & \textbf{-0.76} (<0.01) & -0.68 (0.06) \\
            $\bar{S}$ & \textbf{0.64} (0.01) & \textbf{0.70} (<0.01) & \textbf{0.68} (0.02) & \textbf{0.73} (0.04) \\
            $H_{hr}$ & \textbf{0.72} (<0.01) & 0.43 (0.09) & 0.49 (0.11) & \textbf{0.88} (<0.01) \\
            \midrule

            \multicolumn{5}{c}{\textbf{SF-36 Physical Functioning Scale}} \\
            \midrule
            \textbf{Feature} & \textbf{Event 1} & \textbf{Event 2} & \textbf{Event 3} & \textbf{Extra} \\
            & \small{(n=15)} & \small{(n=16)} & \small{(n=12)} & \small{(n=8)} \\
            \midrule
            $H_s$ & \textbf{0.80} (<0.01) & \textbf{0.78} (<0.01) & \textbf{0.86} (<0.01) & \textbf{0.90} (<0.01) \\
            $R_{HR/S}$ & \textbf{-0.66} (<0.01) & \textbf{-0.76} (<0.01) & \textbf{-0.85} (<0.01) & \textbf{-0.72} (0.04) \\
            $\bar{S}$ & \textbf{0.68} (<0.01) & \textbf{0.69} (<0.01) & \textbf{0.77} (<0.01) & 0.70 (0.05) \\
            $H_{hr}$ & \textbf{0.78} (<0.01) & \textbf{0.53} (0.04) & \textbf{0.61} (0.04) & \textbf{0.85} (<0.01) \\
            \bottomrule
        \end{tabular}
    \end{table}

    We see that steps entropy shows a strong positive correlation with both
    questionnaires, indicating that BCPs with more variable and less predictable
    step patterns tend to report higher physical functioning. Similarly, average
    daily steps showed a strong positive correlation, suggesting that overall
    activity volume is linked to better self-reported function in BCPs. In
    contrast, the heart rate to speed ratio exhibited a strong negative
    correlation, implying that patients who require more cardiovascular effort per
    unit of walking speed report lower physical functioning. Heart rate entropy
    also demonstrated a moderate positive correlation, indicating that greater
    variability may be indicative of better physical functioning. Interestingly,
    none of the features were significantly correlated with physical functioning
    levels in the healthy control group, suggesting that their association with
    physical functioning is specific to the context of breast cancer recovery.

    Overall, our results indicate that the proposed features capture meaningful
    aspects of BCP physical functioning and could serve as an objective measure of
    their functional status. However, despite the strong correlations, the
    relatively small sample size of the BCP group used in this analysis (total of 51
    observations from 23 patients) prevents drawing definitive conclusions.
    Future work will focus on (i) the development of predictive models --- such as
    multivariate regression or machine learning frameworks --- capable of producing
    continuous PF score estimates from wearable data while revealing the
    complementary contribution of individual features, (ii) validation in larger,
    more diverse BCP cohorts and across multiple clinical centers to assess
    generalizability across different patient populations,
    and (iii) systematic investigation of the sensitivity of the extracted features
    to observation window size, with the goal of determining the minimum viable
    monitoring period for reliable PF estimation.

    \section{Conclusion}
    In this paper, we introduced a novel set of features extracted from heart rate and step count
    time series and examined their association with the physical
    functioning status of breast cancer patients and healthy controls, using
    real-world data collected during the REBECCA-CRF study.
    We demonstrated strong and consistent correlations
    between several of the proposed features and self-reported PF scores across multiple
    clinical visits, establishing the feasibility of objectively and
    continuously monitoring functional status without relying solely on
    self-reporting. To that end, our work paves the way for the development of novel digital
    PF biomarkers and prediction models in the context of breast cancer management.

    \bibliographystyle{IEEEtran}
    \bibliography{root}

@article{treanor_methodological_2015,
	title = {A methodological review of the {Short} {Form} {Health} {Survey} 36 ({SF}-36) and its derivatives among breast cancer survivors},
	volume = {24},
	issn = {1573-2649},
	doi = {10.1007/s11136-014-0785-6},
	language = {en},
	number = {2},
	urldate = {2025-07-16},
	journal = {Quality of Life Research},
	author = {Treanor, Charlene and Donnelly, Michael},
	month = feb,
	year = {2015},
	pages = {339--362},
}

@article{luo_assessment_2020,
	title = {Assessment of {Fatigue} {Using} {Wearable} {Sensors}: {A} {Pilot} {Study}},
	volume = {4},
	issn = {2504-110X},
	shorttitle = {Assessment of {Fatigue} {Using} {Wearable} {Sensors}},
	doi = {10.1159/000512166},
	language = {eng},
	number = {Suppl 1},
	journal = {Digital Biomarkers},
	author = {Luo, Hongyu and others},
	year = {2020},
	pmid = {33442581},
	pmcid = {PMC7768149},
	pages = {59--72},
}

@article{paraschiv-ionescu_unraveling_2013,
	title = {Unraveling dynamics of human physical activity patterns in chronic pain conditions},
	volume = {3},
	copyright = {2013 The Author(s)},
	issn = {2045-2322},
	doi = {10.1038/srep02019},
	language = {en},
	number = {1},
	urldate = {2025-07-16},
	journal = {Scientific Reports},
	author = {Paraschiv-Ionescu, Anisoara and Buchser, Eric and Aminian, Kamiar},
	month = jun,
	year = {2013},
	note = {Publisher: Nature Publishing Group},
	pages = {2019},
}

@article{fayers_quality_2002,
	title = {Quality of life research within the {EORTC}—the {EORTC} {QLQ}-{C30}},
	volume = {38},
	issn = {0959-8049},
	doi = {10.1016/S0959-8049(01)00448-8},
	urldate = {2025-07-16},
	journal = {European Journal of Cancer},
	author = {Fayers, P and Bottomley, A},
	month = mar,
	year = {2002},
	pages = {125--133},
}

@article{zomkowski_physical_2018,
	title = {Physical symptoms and working performance in female breast cancer survivors: a systematic review},
	volume = {40},
	issn = {0963-8288},
	shorttitle = {Physical symptoms and working performance in female breast cancer survivors},
	doi = {10.1080/09638288.2017.1300950},
	number = {13},
	urldate = {2025-07-16},
	journal = {Disability and Rehabilitation},
	author = {Zomkowski, Kamilla and others},
	month = jun,
	year = {2018},
	pmid = {28325132},
	note = {Publisher: Taylor \& Francis
\_eprint: https://doi.org/10.1080/09638288.2017.1300950},
	pages = {1485--1493},
}

@article{mu_intelligent_2024,
	title = {Intelligent fatigue detection based on hierarchical multi-scale {ECG} representations and {HRV} measures},
	volume = {92},
	issn = {1746-8094},
	doi = {10.1016/j.bspc.2024.106127},
	urldate = {2025-07-16},
	journal = {Biomedical Signal Processing and Control},
	author = {Mu, Siqi and others},
	month = jun,
	year = {2024},
	pages = {106127},
}

@article{gresham_wearable_2018,
	title = {Wearable activity monitors to assess performance status and predict clinical outcomes in advanced cancer patients},
	volume = {1},
	copyright = {2018 The Author(s)},
	issn = {2398-6352},
	doi = {10.1038/s41746-018-0032-6},
	language = {en},
	number = {1},
	urldate = {2025-07-16},
	journal = {npj Digital Medicine},
	author = {Gresham, Gillian and others},
	month = jul,
	year = {2018},
	note = {Publisher: Nature Publishing Group},
	pages = {27},
}

@article{gupta_feasibility_2018,
	title = {Feasibility of {Wearable} {Physical} {Activity} {Monitors} in {Patients} {With} {Cancer}},
	doi = {10.1200/CCI.17.00152},
	number = {2},
	urldate = {2025-07-16},
	journal = {JCO Clinical Cancer Informatics},
	author = {Gupta, Arjun and others},
	month = apr,
	year = {2018},
	note = {Publisher: Wolters Kluwer},
	pages = {1--10},
}

@article{ballinger_personalized_2021,
	title = {A {Personalized}, {Dynamic} {Physical} {Activity} {Intervention} {Is} {Feasible} and {Improves} {Energetic} {Capacity}, {Energy} {Expenditure}, and {Quality} of {Life} in {Breast} {Cancer} {Survivors}},
	volume = {11},
	issn = {2234-943X},
	doi = {10.3389/fonc.2021.626180},
	language = {English},
	urldate = {2025-07-17},
	journal = {Frontiers in Oncology},
	author = {Ballinger, Tarah J. and others},
	month = apr,
	year = {2021},
	note = {Publisher: Frontiers},
}

@article{bender_physical_2021,
	title = {Physical activity, cardiorespiratory fitness, and cognitive function in postmenopausal women with breast cancer},
	volume = {29},
	issn = {1433-7339},
	doi = {10.1007/s00520-020-05865-4},
	language = {en},
	number = {7},
	urldate = {2025-07-17},
	journal = {Supportive Care in Cancer},
	author = {Bender, Catherine M. and others},
	month = jul,
	year = {2021},
	pages = {3743--3752},
}

@misc{bourgos_2025_17378308,
  author       = {Bourgos, Paraskevas and
               others},
  title        = {{REBECCA}: A System for Real World Behavioural Data
                   in Breast Cancer Clinical Research and Patient
                   Care
                  },
  month        = oct,
  year         = 2025,
  publisher    = {Zenodo},
  doi          = {10.11159/icbes25.185},
}

@article{stensland2026monitoring,
  title={Monitoring cancer-related fatigue and quality of life in breast and prostate cancer patients after primary
treatment: a study protocol for the {REBECCA} trials in {N}orway},
  author={Stensland, Marius and others},
  journal={Clinical and Experimental Medicine},
  year={2026},
  publisher={Springer}
}

\end{document}